\documentclass[twocolumn,aps,pra]{revtex4-1}
\usepackage{epsfig}
\usepackage[english]{babel}
\usepackage{latexsym}
\usepackage{subfigure}
\usepackage{graphics}
\usepackage{epstopdf}
\usepackage{dcolumn}
\usepackage{amsmath}
\usepackage{hyperref}
\usepackage{amssymb}
\usepackage{appendix}
\usepackage{color}
\usepackage{lineno}
\usepackage{soul}
\usepackage{booktabs}
\usepackage{longtable}
\usepackage{multirow}
\usepackage{array}
\usepackage{ulem}
\usepackage{bm}

\begin{document}

\title{All-optical reconstruction of valley polarization through helicity-resolved high-harmonic generation}

\author{Xiaoyu Bu, Yan Meng, Xiaohui Zhao$^{*}$, Rongxiang Zhang$^{\dagger}$, and Fulong Dong$^{\ddagger}$,}

\date{\today}

\begin{abstract}
We theoretically investigate valley-resolved high-order harmonic generation in gapped graphene driven by elliptically polarized laser fields. 
Using two-band density-matrix simulations and an electron-hole recombination trajectory model, we find that the two inequivalent valleys emit harmonics with opposite helicities. 
Under an elliptically polarized field, these emissions occur predominantly in different half cycles of the laser field. 
Time-dependent density functional theory calculations for monolayer MoS$_2$ show the same temporal separation of harmonic emissions with opposite helicities, supporting the generality of this valley-dependent chiral response. 
We further propose an all-optical scheme to reconstruct valley polarization from helicity-resolved harmonic signals. 
A circularly polarized pulse first prepares a valley population imbalance.
A subsequent elliptically polarized laser induces different changes in the harmonic intensities of opposite helicities through Pauli blocking. 
The ratio of these intensity changes provides a direct measure of the valley polarization. 
Our results demonstrate that chiral high-harmonic emission can serve as an all-optical probe of ultrafast valley-dependent carrier dynamics.

\end{abstract}
\affiliation{College of Physics Science and Technology, Hebei University, Baoding 071002, China}

\maketitle

\section{Introduction}
High-order harmonic generation (HHG) is a fundamental nonlinear optical process that provides access to ultrafast electron dynamics on subcycle and attosecond time scales \cite{ALHuillier,Corkum1,Lewenstein,Ferenc}.
Originally developed in atomic and molecular gases, HHG has been extended to solids, where it offers new opportunities to probe the electronic structure and nonequilibrium dynamics of condensed-matter systems \cite{Ghimire,Luu,CHeide}.
Because harmonic emission is intimately linked to laser-driven carrier motion, solid-state HHG provides a powerful all-optical approach to probing electronic band structures and ultrafast carrier dynamics \cite{Keisuke,NYoshikawa,Dong1,Corkum3,Lanin,Garg}.
Moreover, its sensitivity to the symmetry and geometric properties of electronic states makes HHG particularly well suited for investigating quantum materials with multiple internal electronic degrees of freedom \cite{TTLuu,LYue2,AJUNar}.

Among these degrees of freedom, the valley degree of freedom has attracted considerable interest in low-dimensional materials with hexagonal lattice structures \cite{NYoshikawa2,Rost}. 
Inequivalent valleys provide distinct electronic states in momentum space, and an asymmetric population between them gives rise to valley polarization \cite{FLanger,REFSilva}. 
The ability to generate, manipulate, and detect valley polarization is central to the development of valley-dependent optoelectronic functionalities.
It also provides opportunities for encoding and processing information using the valley degree of freedom \cite{AGindl,ITyulnev}.
Since the two inequivalent valleys can exhibit distinct optical responses, strong-field laser fields offer a natural route for controlling valley-selective carrier dynamics \cite{MYuHDu,WLiXZhu}. 
In this context, HHG is particularly attractive because its ultrafast temporal resolution enables the valley-dependent electronic response to be probed on the time scale of the driving field \cite{PLan}.

Recent studies have shown that inequivalent valleys in hexagonal materials can generate harmonic radiation with distinct polarization properties.
In particular, the $\textrm{K}_1$ and $\textrm{K}_2$ valleys emit harmonics with opposite helicities \cite{SMitra}.
This valley-dependent chiral response provides a potential route to probing valley populations through HHG.
However, extracting valley-resolved information from the total HHG signal remains challenging because the contributions from the two valleys are superimposed in the measured spectrum.
When the two valleys are populated equally, their opposite chiral responses can largely cancel, leading to a weak or even vanishing net harmonic helicity.
Therefore, an important challenge is to establish an all-optical method that directly connects polarization-resolved harmonic signals to the populations in individual valleys, enabling quantitative reconstruction of valley polarization.

In this work, we theoretically investigate valley-resolved HHG in gapped graphene driven by elliptically polarized fields.
Using two-band density-matrix simulations and an electron-hole recombination trajectory model, we show that the two inequivalent valleys emit harmonics with opposite helicities due to their distinct interband-current phase differences, as shown in Fig. 1(a).
Their emissions are also temporally separated within an optical cycle, allowing their contributions to be distinguished.
Time-dependent density functional theory (TDDFT) calculations for monolayer MoS$_2$ further confirm the generality of these valley-dependent chiral responses.
Building on these findings, we propose an all-optical scheme to reconstruct valley polarization.
Pre-excited valley populations modify the subsequent HHG through Pauli blocking, resulting in helicity-dependent changes in the harmonic yield.
The ratio of these intensity changes provides a direct measure of the valley populations, enabling quantitative reconstruction of valley polarization, as shown in Fig. 1(b).
Our results establish a direct link between valley populations and chiral HHG, offering an all-optical approach to probing ultrafast valley dynamics in two-dimensional materials.

\begin{figure}[t]
\begin{center}
\includegraphics[width=8.5cm,height=6cm]{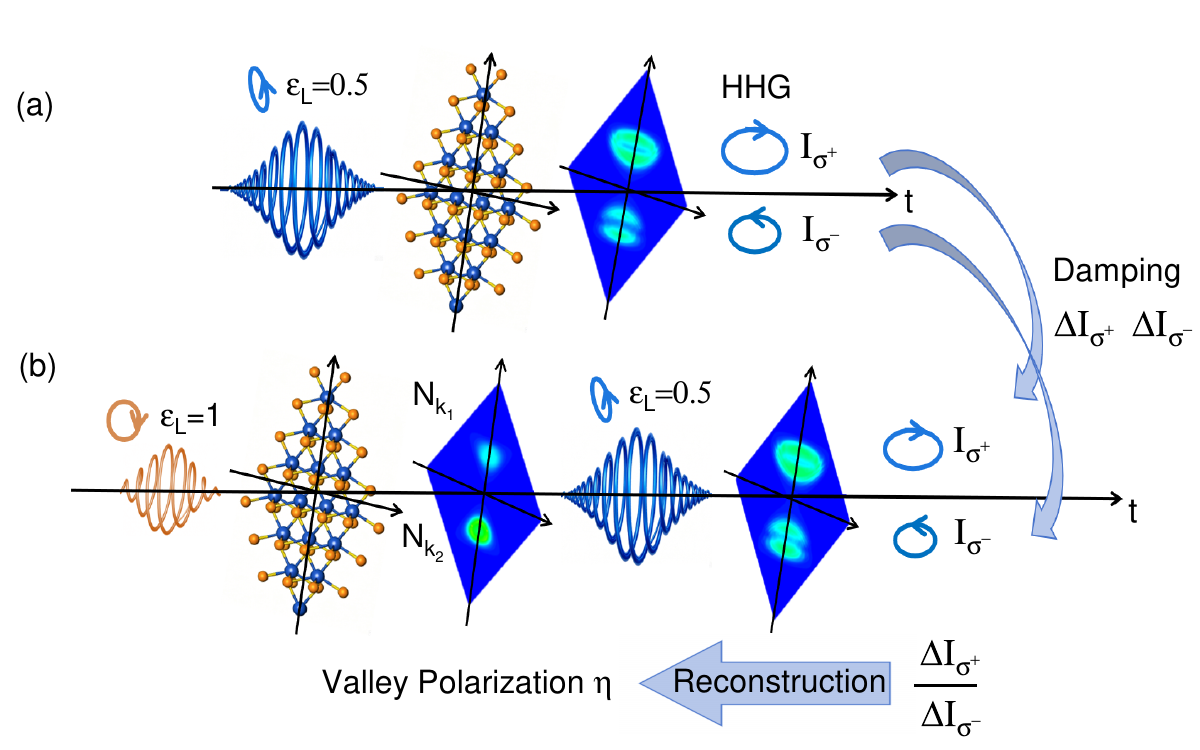}
\caption{
(a) Schematic illustration of HHG in gapped graphene driven by an elliptically polarized laser field with $\varepsilon_L=0.5$. 
The two inequivalent valleys, $\textrm{K}_1$ and $\textrm{K}_2$, generate harmonics with opposite helicities (i.e., right- and left-handed).
(b) Schematic of valley polarization reconstruction.
A circularly polarized pulse ($\varepsilon_L=1$) is first applied to create an imbalance of electron populations between the $\textrm{K}_1$ and $\textrm{K}_2$ valleys.
Subsequently, an elliptically polarized pulse ($\varepsilon_L=0.5$) generates helicity-resolved HHG signals, whose intensity changes $\Delta I_{\sigma^+}$ and $\Delta I_{\sigma^-}$ encode the valley population imbalance and enable the reconstruction of the valley polarization $\eta$. 
The diamond-shaped plots represent the momentum-resolved electron population in the $c$ band.
}
\end{center}
\end{figure}

This paper is organized as follows: We describe our calculation methods for the two-band density-matrix equations (TBDMEs) and present the corresponding simulation results in Sec. \ref{s2}. 
Section \ref{s3} presents our analytical model and reveals the origin of alternating harmonic helicity. 
we propose an all-optical scheme to reconstruct valley polarization in Sec. \ref{s4}.
Section \ref{s5} presents our conclusion. 
Throughout the paper, atomic units are used if not specified.

\section{Numerical calculation methods and results}
\label{s2}

\subsection{Numerical simulation methods based on TBDMEs of gapped graphene}
\label{s2A}
Within the tight-binding approximation and using Bloch states as basis vectors, the Hamiltonian derived from $\pi$ orbitals of gapped graphene can be expressed as
$
H_0=\left(\begin{array}{cc}
\Delta_g/2 & \gamma_0 f(\mathbf{k}) \\
\gamma_0 f^*(\mathbf{k}) & -\Delta_g/2
\end{array}\right).
$
Here $\gamma_0 = 0.1$ a.u. denotes the hopping energy \cite{AHCNeto}, and $f(\textbf{k})=e^{i \texttt{k}_{x}d}+2\cos( \sqrt{3}\texttt{k}_{y}d / 2)e^{-i\texttt{k}_{x}d/2}$ in which $d = 1.42 \text{\AA}$ is the carbon--carbon bond length.
The energy eigenvalues of the conduction $(c)$ and valence $(v)$ bands are $\textrm{E}_c(\mathbf{k})=-\textrm{E}_v(\mathbf{k})= \sqrt{\gamma_0^2 |f(\mathbf{k})|^2 + \Delta_g^{2}/4}$.

We numerically simulate the currents in gapped graphene by solving the two-band density-matrix equations (TBDMEs) in the Houston representation \cite{WVHouston}.
Within the dipole approximation, these equations are

\vspace{-0.4cm}
\begin{align}
i \frac{d}{d t} \rho_{m n}^{\mathbf{K}_{0}}(t) &= \left[\textrm{E}_{mn} (\mathbf{K}_{t}) - i \tilde{\delta}_{mn}/T_2 \right] \rho_{m n}^{\mathbf{K}_{0}}(t)  \nonumber\\
&+ \textit{\textbf{F}}(t) \cdot \sum_{l} \left[\mathbf{D}_{m l}^{\mathbf{K}_{t}} \rho_{l n}^{\mathbf{K}_{0}}(t)-\mathbf{D}_{l n}^{\mathbf{K}_{t}} \rho_{m l}^{\mathbf{K}_{0}}(t)\right],
\end{align}
where $\rho_{mn}^{\mathbf{k}}$ is the density matrix elements.
Here, $m$ and $n$ denote the $v$ or $c$ bands.
$\textrm{E}_{mn} (\mathbf{k}) = \textrm{E}_{m} (\mathbf{k}) - \textrm{E}_{n} (\mathbf{k})$ represents the energy difference between the $m$ and $n$  bands.
$\tilde{\delta}_{mn}=1-\delta_{mn}$, and the dephasing rate is $1/T_2=0.005$ a.u. \cite{YKim} .
The transition dipole elements are given by $\textbf{\textrm{D}}_{mn}^{\textbf{k}} = i \langle u_{m,\textbf{k}}(\textbf{r}) \vert \nabla_{\textbf{k}} \vert u_{n,\textbf{k}}(\textbf{r}) \rangle$, where $u_{m,\textbf{k}}(\textbf{r})$ denotes the periodic part of the Bloch wavefunction for $m$ band of gapped graphene.

In Eq. (1), the crystal quasimomentum is given by $\textbf{K}_{t} = \textbf{K}_{0} + \textit{\textbf{A}}(t)$, where $\mathbf{K}_0 = (\textrm{K}_{0x},\textrm{K}_{0y})$ lies within the first Brillouin zone (BZ).
The vector potential $\textbf{\textit{A}}(t)$ is defined as $\textbf{\textit{A}}(t) = -\int^{t} \mathbf{\textit{F}}(\tau)d\tau$, where $\textbf{\textit{F}}(t)=F_e(t)\textit{\textbf{e}}=\frac{F_0}{\sqrt{1+\varepsilon_{L}^2}}\ f(t)\{ \sin(\omega_0t)\hat{\textit{\textbf{x}}}+\varepsilon_{L}\cos(\omega_0t+\varphi)\hat{\textit{\textbf{y}}}\}$ is the electric field of the laser pulse.
Here, $f(t) = \sin^{2}(\omega_0 t/2n)$ is the laser envelope function with $n=16$.
The frequency $\omega_0$ corresponds to a wavelength of $2500$ nm,
and the amplitude $F_e$ corresponds to an intensity of $1 \times 10^{12}$ W/cm$^2$.
The unit vector $\textit{\textbf{e}}$ indicates the polarization direction of the electric field, and $\varepsilon_{L}$ is the laser ellipticity.

Equation (1) can be solved numerically using the standard fourth-order Runge-Kutta algorithm.
The total harmonic yield is calculated by
\begin{eqnarray}
\begin{aligned}
H_{\textrm{tot}}(\omega)=\omega^{2} (\vert j_{\parallel}(\omega) \vert^{2} + \vert j_{\perp}(\omega) \vert^{2}),
\end{aligned}
\end{eqnarray}
where $\textit{j}_{\mu}(\omega)= \int^{\infty}_{-\infty} \textit{\textbf{e}}_{\mu} \cdot \textbf{\textit{j}}_{\textrm{tot}}(t) e^{-i \omega t} dt$,
and $\mu$ denotes $\parallel$ or $\perp$. 
Here, the total current $\textbf{\textit{j}}_{\textrm{tot}}(t)=\sum_{\textbf{K}_{0}\in{\textrm{BZ}}} \sum_{m,n}\rho_{nm}^{\mathbf{K}_{0}}(t) \mathbf{P}_{m n}^{\mathbf{K}_{t}}$. 
In addition, we define the valley-resolved currents as $\textbf{\textit{j}}_{\textrm{K}_{1}}(t)=\sum_{\mathbf{K}_0}^{\textrm{K}_{0y}>0} \sum_{m,n}\rho_{nm}^{\mathbf{K}_{0}}(t) \mathbf{P}_{m n}^{\mathbf{K}_{t}}$, 
$\textbf{\textit{j}}_{\textrm{K}_{2}}(t)=\sum_{\mathbf{K}_0}^{\textrm{K}_{0y}<0} \sum_{m,n}\rho_{nm}^{\mathbf{K}_{0}}(t) \mathbf{P}_{m n}^{\mathbf{K}_{t}}$, corresponding to the $\textrm{K}_1$ and $\textrm{K}_2$ valleys, respectively.

The time-frequency distribution of the HHG can be evaluated by
\begin{eqnarray}
\begin{aligned}
H^{\textrm{tf}}(\omega,t) = \omega^2 (\vert \textit{j}^{\textrm{tf}}_{\parallel}(\omega,t) \vert^2 + \vert \textit{j}^{\textrm{tf}}_{\perp}(\omega,t) \vert^2),
\end{aligned}
\end{eqnarray}
in which $\textit{j}^{\textrm{tf}}_{\mu}(\omega,t) = \int_{t-T}^{t+T} d t^{\prime} \textit{\textbf{e}}_{\mu} \cdot \textbf{\textit{j}}(t^{\prime}) W(t^{\prime} - t) e^{i \omega t^{\prime}}$, here $W(x) = \dfrac{1}{\sqrt{2 \pi \tau_0}} e^{-x^2 / 2\tau_0^2}$ with $\tau_0 = 1/3 \omega_0$.
$T = 2 \pi / \omega_0$ is the laser period.

The harmonic ellipticity can be evaluated using the following relation \cite{zhushiyi}:
\begin{eqnarray}
\begin{aligned}
\varepsilon_H = \textrm{sgn}
 [\sin(\delta)]\sqrt{\dfrac{1+r^{2}-\sqrt{1+2r^{2}\cos 2\delta + r^{4}}}{1+r^{2}+\sqrt{1+2r^{2} \cos 2\delta + r^{4}}}},
\end{aligned}
\end{eqnarray}
where $r(\omega,t)=\vert \textit{j}^{\textrm{tf}}_{\perp}(\omega,t) \vert / \vert \textit{j}^{\textrm{tf}}_{\parallel}(\omega,t) \vert$ is the amplitude ratio between the perpendicular and parallel components.
The phase difference is given by $\delta(\omega,t) = \varphi_{\parallel}(\omega,t) - \varphi_{\perp}(\omega,t) + q \pi$, constrained to the range of $[-\pi/2,\pi/2]$, where $\varphi_{\mu}(\omega,t) =$ arg$[\textit{j}^{\textrm{tf}}_{\mu}(\omega,t)]$ and $q$ is an integer.

\subsection{Time-dependent density-functional theory}
\label{s2B}

To verify our primary high-harmonic generation phenomena, we performed numerical simulations using time-dependent density-functional theory  (TDDFT) \cite{Ullrich}.
We select $\textrm{MoS}_2$ as the target material, whose energy gap is close to $0.07$ a.u. \cite{KFMak}.
In our simulation, we use the norm-conserving pseudopotentials
and exchange-correlation potential based on the generalized gradient approximation in the Perdew-Burke-Ernzerhof parametrization \cite{JohnPPerdew}. 
Within the TDDFT framework, the wave function evolution is computed by propagating the Kohn-Sham equations \cite{MALMarques,GuillaumeLB,ZNourbakhsh}.
In the calculation of the harmonic generation in monolayer $\textrm{MoS}_2$, a $40 \times 40 \times 1$ k-point mesh is used to sample the first BZ and the real-space spacing is $0.2 \text{\AA}$. 

The vector potential $\textbf{A}(t)$ is the same as that used in TBDMEs. 
We compute the total electronic current j(r,t) from time-evolved wave functions. 
The harmonic yield can be evaluated using $H_{\textrm{tot}}(\omega) = \omega^{2} (\vert j_{\parallel}(\omega) \vert^{2} + \vert j_{\perp}(\omega) \vert^{2})$, in which $j_{\parallel(\perp)}(\omega) = \mathbf{T}_F[\textit{\textbf{e}}_{\parallel(\perp)} \cdot \int_\Omega d^3\mathbf{r}\mathbf{j}(\mathbf{r},t)]$, where $\Omega$ is the volume of the physical system. The OCTOPUS package \cite{Strubbe,Strubbe2} is employed to perform these simulations.

\subsection{Numerical simulation results}
\label{s2C}

\begin{figure}[t]
\begin{center}
\includegraphics[width=8.5cm,height=11cm]{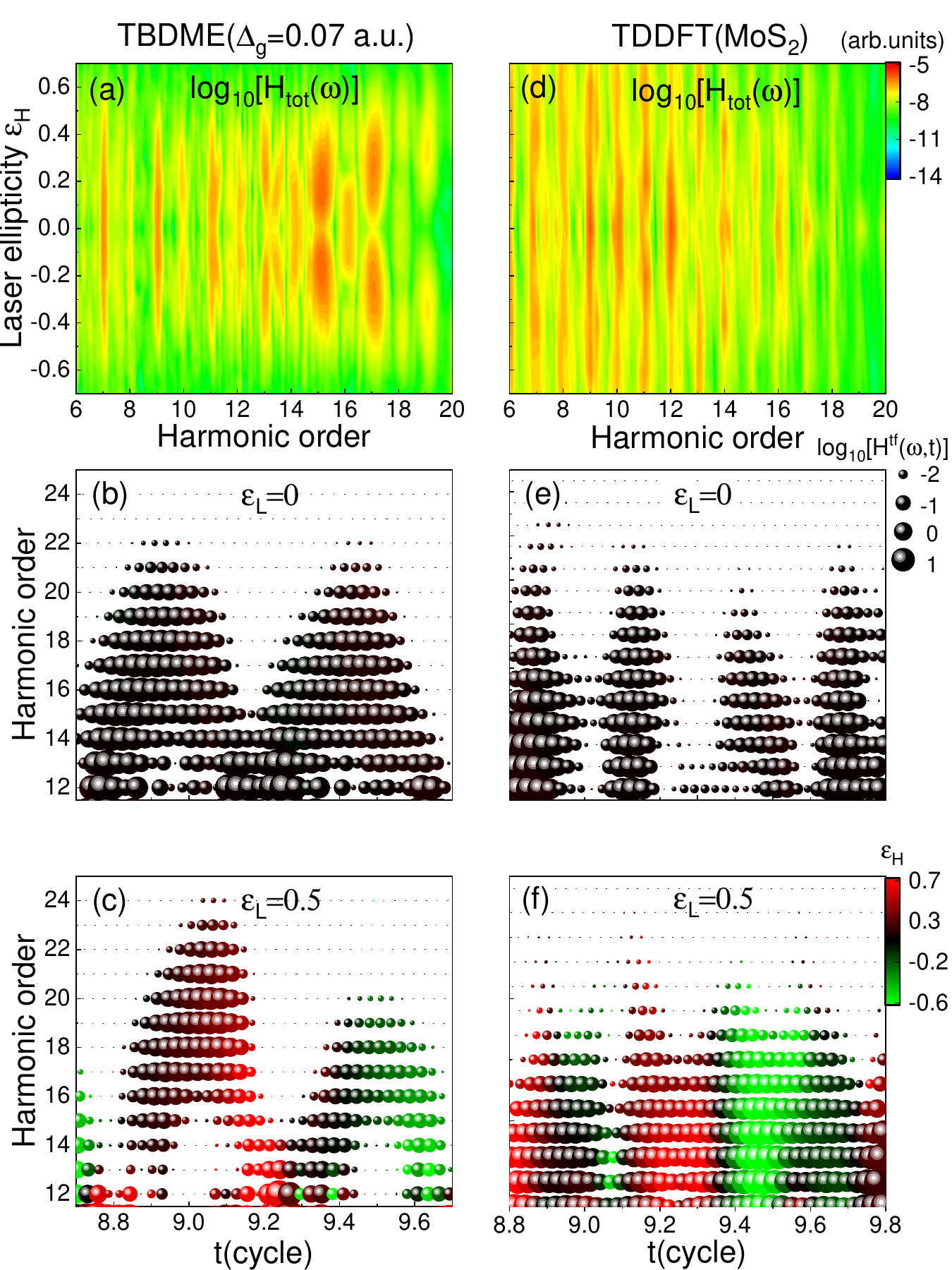}
\caption{
Comparison between the TBDME and TDDFT results.
(a) Dependence of the total harmonic intensity on the driving-field ellipticity for gapped graphene with $\Delta_g=0.07$ a.u., calculated using the TBDME framework.
(b,c) Time-frequency distributions of the harmonic emission under linearly polarized ($\varepsilon_L=0$) and elliptically polarized ($\varepsilon_L=0.5$) driving fields, respectively.
The sphere size represents the harmonic emission intensity, while its color indicates the harmonic ellipticity.
(d–f) Corresponding results obtained from TDDFT simulations of monolayer $\mathrm{MoS}_2$.
}
\end{center}
\end{figure}

Figure 2(a) shows the dependence of the total harmonic intensity on the driving-field ellipticity for gapped graphene with $\Delta_g=0.07$ a.u., calculated using Eq. (2). 
Within the ellipticity range of $-0.5\leq\varepsilon_L\leq0.5$, pronounced harmonic emission is observed. 
Both odd- and even-order harmonics are clearly resolved, with the harmonic cutoff extending to the $19^{\textrm{th}}$ order.

To resolve the subcycle dynamics of the harmonic emission, Figs. 2(b) and 2(c) show the time-frequency distributions of the total harmonic emission.
The sphere size represents the instantaneous harmonic intensity calculated from Eq. (3), while the color indicates the harmonic ellipticity evaluated using Eq. (4).
Here, red and green denote right- and left-handed harmonic emission, respectively.
For a linearly polarized driving field ($\varepsilon_L=0$), as shown in Fig. 2(b), all harmonics exhibit zero ellipticity.
By contrast, for an elliptically polarized field with $\varepsilon_L=0.5$ [Fig. 2(c)], the helicity of the harmonic emission becomes strongly dependent on the subcycle dynamics.
Right-handed harmonics are predominantly emitted during the first half cycle, whereas left-handed harmonics are mainly generated during the second half cycle.

To verify that these features are not specific to the TBDME simulations of gapped graphene, we further perform TDDFT simulations for monolayer $\textrm{MoS}_2$. 
As shown in Fig. 2(d), both odd- and even-order harmonics are clearly resolved, with the harmonic cutoff extending to the $19^{\textrm{th}}$ order, consistent with the results for gapped graphene.
The corresponding time-frequency distribution under a linearly polarized driving field is shown in Fig. 2(e), where the instantaneous ellipticity remains zero throughout the laser cycle, consistent with Fig. 2(b). 
When the driving-field ellipticity is increased to $\varepsilon_L=0.5$, the time-frequency distribution in Fig. 2(f) exhibits the same subcycle helicity dependence observed in Fig. 2(c).
The right-handed harmonics dominate during the first half cycle, whereas left-handed harmonics prevail during the second half cycle. 
These TDDFT results thus corroborate the main features revealed by the TBDME simulations.

\begin{figure}[t]
\begin{center}
\includegraphics[width=8.5cm,height=8cm]{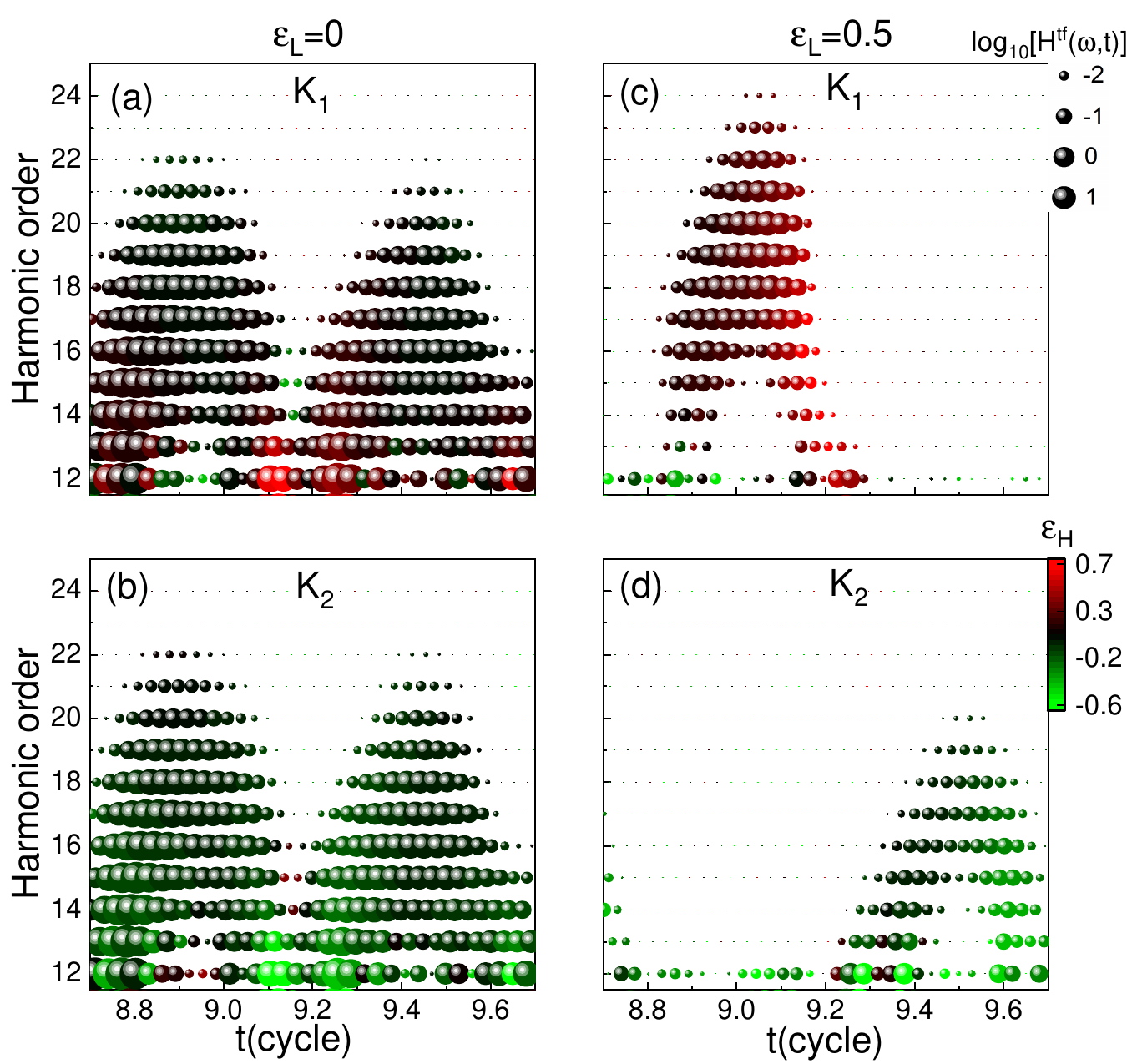}
\caption{
Valley-resolved time-frequency distributions of harmonic emission from the $\textrm{K}_1$ [(a), (c)] and $\textrm{K}_2$ valleys [(b), (d)] under linearly polarized ($\varepsilon_L=0$) [(a), (c)] and elliptically polarized ($\varepsilon_L=0.5$) [(b), (d)] driving fields.
The size of spheres represents the harmonic emission intensity, while their color indicates the harmonic ellipticity $\varepsilon_H$.
}
\end{center}
\end{figure}

Next, we analyze the individual harmonic emissions from the $\textrm{K}_1$ and $\textrm{K}_2$ valleys separately. 
Figures 3(a) and 3(b) show the time-frequency distributions of the valley-resolved harmonic emissions from the $\textrm{K}_1$ and $\textrm{K}_2$ valleys, respectively, driven by a linearly polarized laser field.
As shown in Fig. 3(a), the $\textrm{K}_1$ valley exclusively generates right-handed harmonics, whose intensities are slightly higher than those in Fig. 2(b). 
In contrast, the $\textrm{K}_2$ valley exclusively generates left-handed harmonics, as shown in Fig. 3(b), whose intensities are also slightly higher than those Fig. 2(b).
These results demonstrate that the linearly polarized harmonics originates from the coherent superposition of two valley-selective emissions with opposite helicities.

Figures 3(c) and 3(d) show the time-frequency distributions of the harmonic emissions from the $\textrm{K}_1$ and $\textrm{K}_2$ valleys, respectively, driven by an elliptically polarized field with $\varepsilon_L=0.5$.
The elliptical field induces a pronounced temporal asymmetry in the valley-resolved HHG.
The $\textrm{K}_1$ ($\textrm{K}_2$) valley dominates during the first (second) half optical cycle and generates intense right-handed (left-handed) harmonics.
The combined valley contributions can reproduce the total time-frequency distribution shown in Fig. 2(c).
(See Appendix A for other energy gaps).

\section{Analytical model and mechanism discussion}
\label{s3}
\subsection{Electron-hole recombination trajectory model}

To elucidate the underlying mechanism, we investigate the recombination trajectory model of electron-hole pairs within the framework of the two-band density matrix equation.
Under the strong-field approximation, intraband currents can be neglected, while the interband current plays a dominant role in the process of HHG.
In this approximation, the Fourier transform of the interband current is
\begin{align}
\textit{j}_{\mu}(\omega) \sim & \int d \textbf{K}_{0} \int_{- \infty}^{\infty} d t \int_{-\infty}^{t} dt^{\prime} g_{\mu}(\textbf{K}_{0},t^{\prime},t) e^{-iS_{\mu}(\textbf{K}_{0},t^{\prime},t,\omega)},
\end{align}
in which
$g_{\mu}(\textbf{K}_{0},t^{\prime},t) = -\textit{F}(t^{\prime}) \left| \textrm{D}_{cv,\parallel}^{\mathbf{K}_{t^{\prime}}} \right|  \textrm{E}_{c v} (\textbf{K}_{t}) \left| \textrm{D}_{cv,\mu}^{\mathbf{K}_{t}} \right|$ is a slowly varying term.
Here, $\textrm{D}_{c v,\parallel}^{\textbf{k}} = \textrm{D}_{c v,x}^{\textbf{k}} \cos \theta + \textrm{D}_{c v,y}^{\textbf{k}} \sin \theta$ and $\textrm{D}_{c v,\perp}^{\textbf{k}} = -\textrm{D}_{c v,x}^{\textbf{k}} \sin \theta + \textrm{D}_{c v,y}^{\textbf{k}} \cos \theta$, in which $\textrm{D}_{c v,x(y)}^{\textbf{k}}$ denotes the $x$ ($y$) component of $\textbf{D}_{c v}^{\textbf{k}}$.
In Eq. (5), the semiclassical action is
\begin{align}
S_{\mu} (\textbf{K}_{0},t^{\prime},t,\omega) = & \int_{t^{\prime}}^t  \left[ \textrm{E}_{c v} (\textbf{K}_{\tau}) + \boldsymbol{F}(\tau) \cdot \boldsymbol{\mathcal{A}}(\textbf{K}_{\tau}) \right] d \tau \nonumber \\
& + \alpha_{\mu}(\textbf{K}_{t}) - \alpha_{\parallel}(\textbf{K}_{t^{\prime}})  - \omega t,
\end{align}
in which $\boldsymbol{\mathcal{A}}(\textbf{k}) = \textbf{D}_{c c}^{\textbf{k}} -\textbf{D}_{v v}^{\textbf{k}}$, and
$\alpha_{\mu}(\textbf{k}) = \textrm{arg}(\textrm{D}_{c v,\mu}^{\textbf{k}})$ is the transition dipole phase.

\begin{figure}[t]
\begin{center}
\includegraphics[width=8.5cm,height=8.5cm]{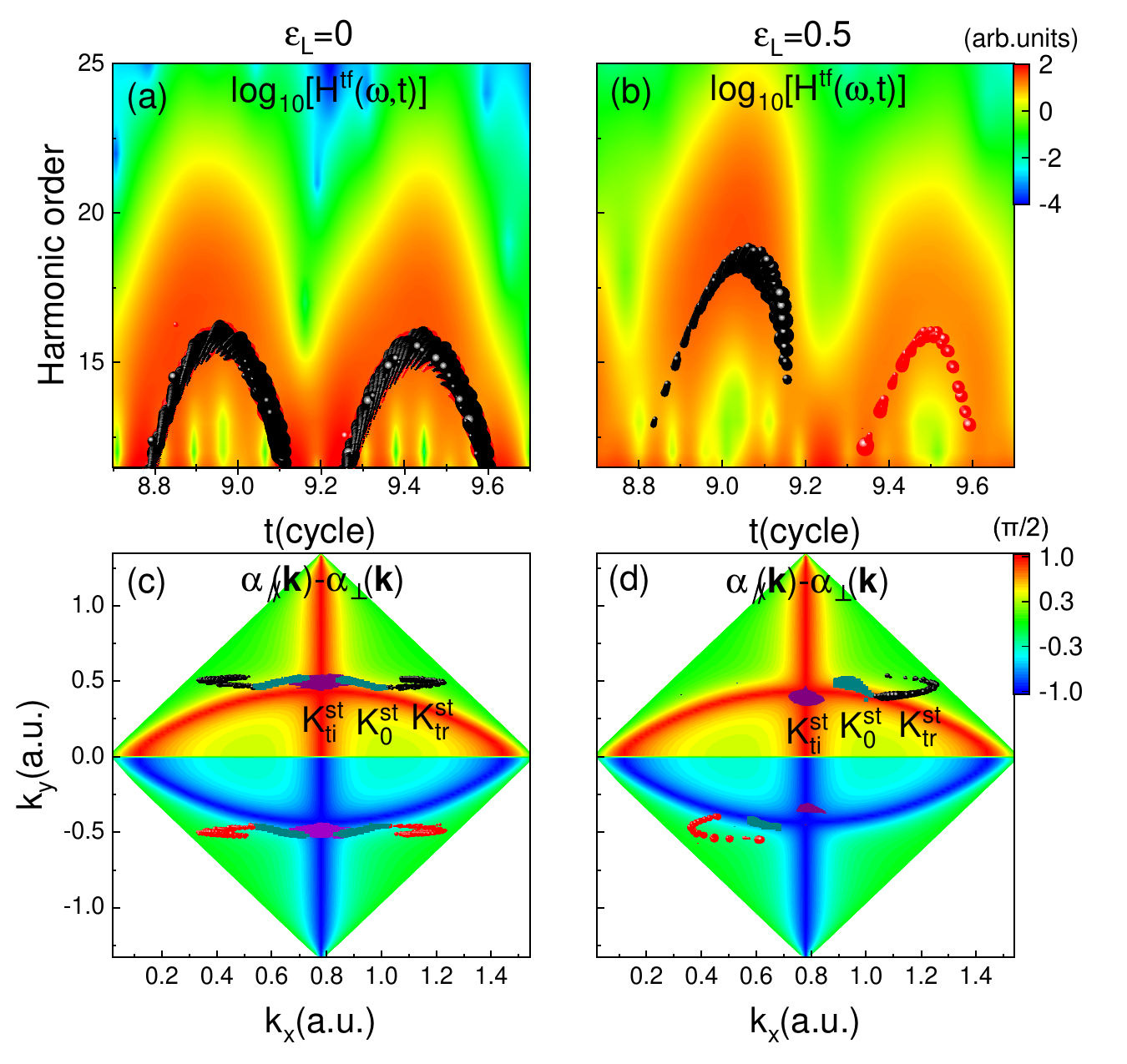}
\caption{
(a) Time-frequency distribution of the total harmonic spectrum driven by a linearly polarized laser (i.e., $\varepsilon_L=0$). 
The overlaid points represent the semiclassical electron–hole recombination trajectories.
The black and red points correspond to the $\textrm{K}_{1}$ and $\textrm{K}_{2}$ valleys, respectively.
(b) Momentum-space distribution of the phase difference $\alpha_{\parallel}(\textbf{k}) - \alpha_{\perp}(\textbf{k})$. 
The saddle-point momenta associated with the electron–hole trajectories are marked by the corresponding colored points. 
(c), (d) Same as (a) and (b), respectively, but for an elliptically polarized driving field with $\varepsilon_L=0.5$.
}
\end{center}
\end{figure}

Next, we apply the steady-phase approximation to the variables $\textbf{\textrm{K}}_0, t, t^{\prime}$, yielding the following equations:

\vspace{-0.4cm}
\begin{subequations}
\begin{align}
\textrm{E}_{c v}(\mathbf{K}^{\textrm{st}}_{t_{i}}) + & \boldsymbol{F}(t_{i}) \cdot \left( \boldsymbol{\mathcal{A}}(\textbf{K}^{\textrm{st}}_{t_{i}})-\nabla_{\mathbf{K}^{\textrm{st}}_{t_{i}}} \alpha_{\|}(\mathbf{K}^{\textrm{st}}_{t_{i}}) \right) \leq \textrm{E}_{i}, \\
\bigg|  \int_{t_i}^{t_r} \nabla_{\mathbf{K}^{\textrm{st}}_{\tau}} & \left( \textrm{E}_{c v}(\mathbf{K}^{\textrm{st}}_{\tau}) + \boldsymbol{F}(\tau) \cdot \boldsymbol{\mathcal{A}}(\mathbf{K}^{\textrm{st}}_{\tau}) \right)   d \tau \nonumber \\
+ & \nabla_{\mathbf{K}^{\textrm{st}}_{t_r}}  \alpha_{\mu}(\mathbf{K}^{\textrm{st}}_{t_r}) - \nabla_{\mathbf{K}^{\textrm{st}}_{t_i}} \alpha_{\parallel}(\mathbf{K}^{\textrm{st}}_{t_i}) \bigg|  =0 , \\
\textrm{E}_{c v}(\mathbf{K}^{\textrm{st}}_{t_r}) + & \boldsymbol{F}(t_r) \cdot\left(\boldsymbol{\mathcal{A}}(\mathbf{K}^{\textrm{st}}_{t_r})-\nabla_{\mathbf{K}^{\textrm{st}}_{t_r}} \alpha_{\mu}(\mathbf{K}^{\textrm{st}}_{t_r})\right)=\omega.
\end{align}
\end{subequations}
In Eqs. (7), $t_i$ and $t_r$ represent the ionization and recombination times of electron-hole pairs, respectively.
Here, $\mathbf{K}^{\textrm{st}}_t = \mathbf{K}_0^{\textrm{st}} + \textit{\textbf{A}}(t)$, where $\mathbf{K}_0^{\textrm{st}} = (\textrm{K}_{0x}^{\textrm{st}},\textrm{K}_{0y}^{\textrm{st}})$ is the saddle-point momentum.

In the process of solving the equation, we first sample the lattice momenta $\textbf{\textrm{K}}_0$ from the first BZ, which correspond to inequivalent electrons.
Next, the electrons oscillate in the reciprocal space driven by the laser.
At time $t_{i}$, determined by the condition in Eq. (7a), electrons are excited from the $v$ to $c$ band, leaving a hole in the $v$ band.
It is important to note that electron excitation can occur not only at the point of minimum energy gap (i.e., the $\textrm{K}$ points), but also in its vicinity \cite{Dong5,MKolesik,LunYue}. 
In our simulation, we set the excitation energy threshold $\textrm{E}_{i}$ to $0.1$ a.u. for gapped graphene of $\Delta_g = 0.07$ a.u. 
Next, the electron-hole pairs are assumed to move in the two-dimensional coordinate space. 
When they recombine, which implies Eq. (7b) is satisfied, the recombination time $t_r$ is obtained. 
Finally, the collision energy $\omega$ can be calculated using Eq. (7c).

Under the steady-phase approximation, for a specific saddle-point momentum $\textbf{K}_{0}^{\textrm{st}}$, Eq. (6) can be integrated to
$\textit{j}_{\mu}^{\textbf{K}_{0}^{\textrm{st}}}(\omega) \varpropto  g_{\mu}(\textbf{K}_{0}^{\textrm{st}},t_i,t_r) \dfrac{e^{-iS_{\mu}(\textbf{K}_{0}^{\textrm{st}},t_i,t_r,\omega)}}{\sqrt{\vert \operatorname{det} [S_{\mu}^{\prime \prime}(\textbf{K}_{0}^{\textrm{st}},t_i,t_r,\omega)] \vert}}$,
where $S_{\mu}^{\prime \prime}(\textbf{K}_{0}^{\textrm{st}},t_i,t_r,\omega)$ is the Hessian matrix \cite{Dong5}.
Therefore, for a specific saddle-point trajectory, the phase difference $\delta$ between $\textit{j}_{\parallel}^{\textbf{K}_{0}^{\textrm{st}}}(\omega)$ and $\textit{j}_{\perp}^{\textbf{K}_{0}^{\textrm{st}}}(\omega)$ can be evaluated by
\begin{align}
\delta (\textbf{K}_{0}^{\textrm{st}}) = S_{\parallel}(\textbf{K}_{0}^{\textrm{st}}) - S_{\perp}(\textbf{K}_{0}^{\textrm{st}}) = \alpha_{\parallel}(\mathbf{K}^{\textrm{st}}_{t_r})-\alpha_{\perp}(\mathbf{K}^{\textrm{st}}_{t_r}).
\end{align}

\subsection{Mechanism discuss}
\label{s3B}

Figure 4(a) shows the time-frequency distribution of the total harmonic spectrum for $\varepsilon_L=0$, which is identical to that presented in Fig. 2(b).
The recombination energy $\omega(t_r)$ at the recombination time $t_r$ solved from Eqs. (7) is overlaid in Fig. 4(a), where overlapped black and red points mark the saddle-point trajectory solutions from the $\textrm{K}_{1}$ and $\textrm{K}_{2}$ valleys, respectively.
The saddle-point trajectories qualitatively align with the peak regions of harmonic intensity in the contour map, which verifies that the analytical saddle-point solutions can qualitatively reproduce the full time-frequency evolution of harmonics obtained from numerical simulations.

In Fig. 4(b), the contour map shows the momentum-space distribution of the phase difference $\alpha_{\parallel}(\mathbf{k})-\alpha_{\perp}(\mathbf{k})$, where the saddle-point momenta $\mathbf{K}_0^{\textrm{st}}$, $\mathbf{K}_{t_i}^{\textrm{st}}$, and $\mathbf{K}_{t_r}^{\textrm{st}}$ are marked by the overlaid points.
According to Eq. (8), the phase difference $\delta = \alpha_{\parallel}(\mathbf{K}^{\textrm{st}}_{t_r})-\alpha_{\perp}(\mathbf{K}^{\textrm{st}}_{t_r})$ is solely determined by the recombination momentum $\mathbf{K}^{\textrm{st}}_{t_r}$.
As shown in Fig. 4(b), the phase difference is positive (negative) for the $\textrm{K}_1$ ($\textrm{K}_2$) valley, leading to right-handed (left-handed) harmonic emission, as observed in Fig. 3(a) [Fig. 3(b)]. 
In addition, for each of the $\textrm{K}_1$ and $\textrm{K}_2$ valleys, there are two trajectory branches within one optical cycle, corresponding to the two branches of the time-frequency distribution observed in Figs. 3(a), 3(b), and 4(a).

As the laser elliptically increases to $\varepsilon_L=0.5$, Fig. 4(c) shows that the recombination trajectories associated with the $\textrm{K}_{1}$ and $\textrm{K}_{2}$ valleys become temporally separated within one optical cycle, resulting in valley-selective harmonic emissions, consistent with the results shown in Figs. 3(c) and 3(d).
Compared with the linearly polarized case, the saddle-point momenta in Fig. 4(d) indicate that electrons in the $\textrm{K}_1$ ($\textrm{K}_2$) valley can recombine with holes predominantly during the first (second) half cycle, emitting right-handed (left-handed) harmonics.

\section{RECONSTRUCTION OF VALLEY POLARIZATION }
\label{s4}
\subsection{Dependence of harmonic intensity on the valley polarization}

Previous studies have shown that an elliptically polarized driving field with $\varepsilon_L=0.5$ can induce valley-selective harmonic emission with opposite helicities in different half-cycles. 
Building on this property, we propose an all-optical scheme for reconstructing valley polarization. 
When the conduction band is initially populated, Pauli blocking suppresses subsequent laser-driven transitions from the $v$ band to the $c$ band, thereby reducing the harmonic emission. 
In a valley-polarized system, where the $c$-band populations differ between the $\textrm{K}_1$ and $\textrm{K}_2$ valleys, this Pauli-blocking effect is valley dependent. 
Consequently, under an elliptically polarized driving field with $\varepsilon_L=0.5$, the valley-selective harmonics with opposite helicities experience different degrees of intensity suppression compared with the case of an initially unpopulated conduction band.
The ratio between these helicity-dependent intensity reductions can therefore be used to reconstruct the magnitude of the valley polarization.

To verify our scheme, the model is first irradiated by a right-handed circularly polarized laser to induce an initial valley polarization, as shown in Fig. 1(b).
The pulse has a sine-squared envelope and a duration of eight optical cycles.
The laser wavelength is $3000$ nm, and its intensity is $2\times10^{11}$ W/cm$^2$.
Sixty femtoseconds after the end of the circularly polarized pulse, an elliptically polarized laser with $\varepsilon_L=0.5$, identical to that introduced in the preceding sections, is applied.

Figure 5(a) displays the time-frequency distribution generated by the elliptically polarized laser field following the circularly polarized pre-excitation.
A clear difference is observed compared with the result shown in Fig. 2(c), which can be attributed to Pauli-blocking effect.
To quantify this difference, we subtract the time-frequency current $\textbf{\textit{j}}_{\circlearrowleft}^{\textrm{tf}}(\omega,t)$ shown in Fig. 5(a) from $\textbf{\textit{j}}^{\textrm{tf}}(\omega,t)$ in Fig. 2(c).
The intensity difference $\Delta H^{\textrm{tf}} (\omega,t) = \omega^2 \left| \textbf{\textit{j}}^{\textrm{tf}}(\omega,t) - \textbf{\textit{j}}_{\circlearrowleft}^{\textrm{tf}}(\omega,t) \right|^2 $ is shown in Fig. 5(b).
The result indicates that compared with Fig. 2(c), the harmonic yields of both the right- and left-handed polarized components in Fig. 5(a) are reduced.
More importantly, the reduction in the right-handed polarized component is smaller than that in the left-handed polarized component.
This difference can be attributed to the unequal $c$-band populations in the two valleys, with the population in the $\textrm{K}_1$ valley being lower than that in the $\textrm{K}_2$ valley, as shown in the inset of Fig. 5.

\begin{figure}[t]
\begin{center}
\includegraphics[width=8.5cm,height=4cm]{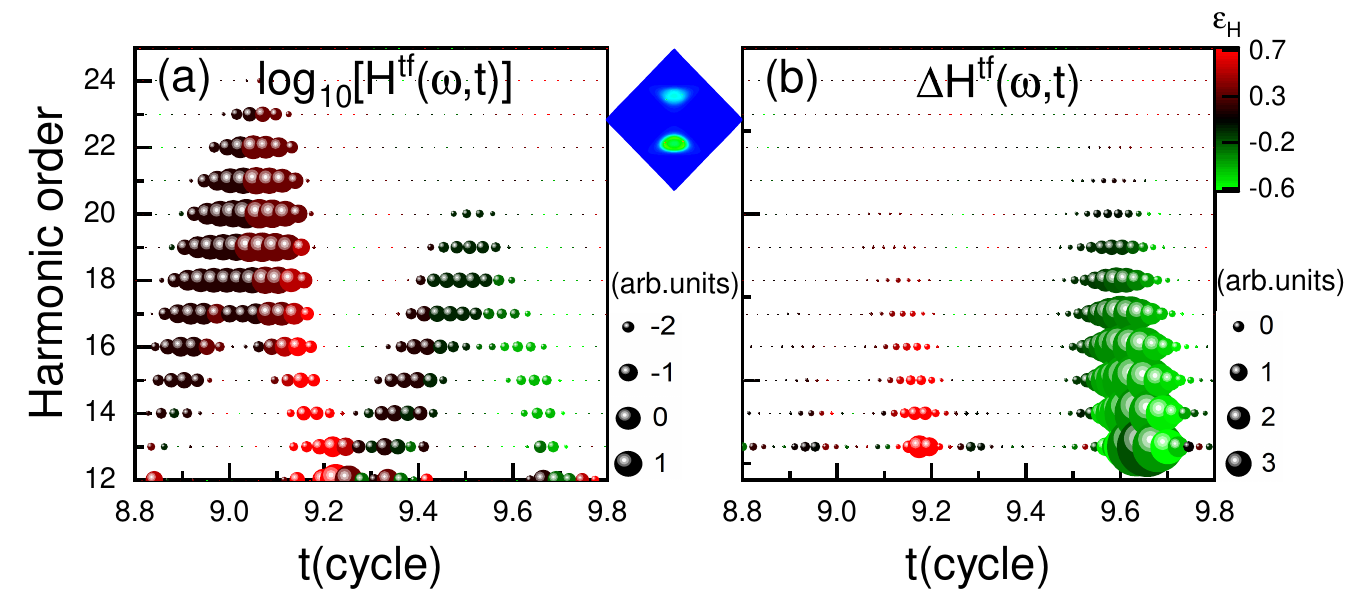}
\caption{
(a) Time-frequency distribution of the harmonic emission generated by the combined circularly polarized and elliptically polarized fields as shown in Fig. 1(b).
The inset shows the momentum-resolved electron population of the $c$ band induced by the circularly polarized pulse.
(b) Harmonic intensity difference obtained by subtracting the harmonic signals generated by the combined fields from that generated by the elliptically polarized field alone. 
}
\end{center}
\end{figure}

\begin{figure}[t]
\begin{center}
\includegraphics[width=8.5cm,height=8cm]{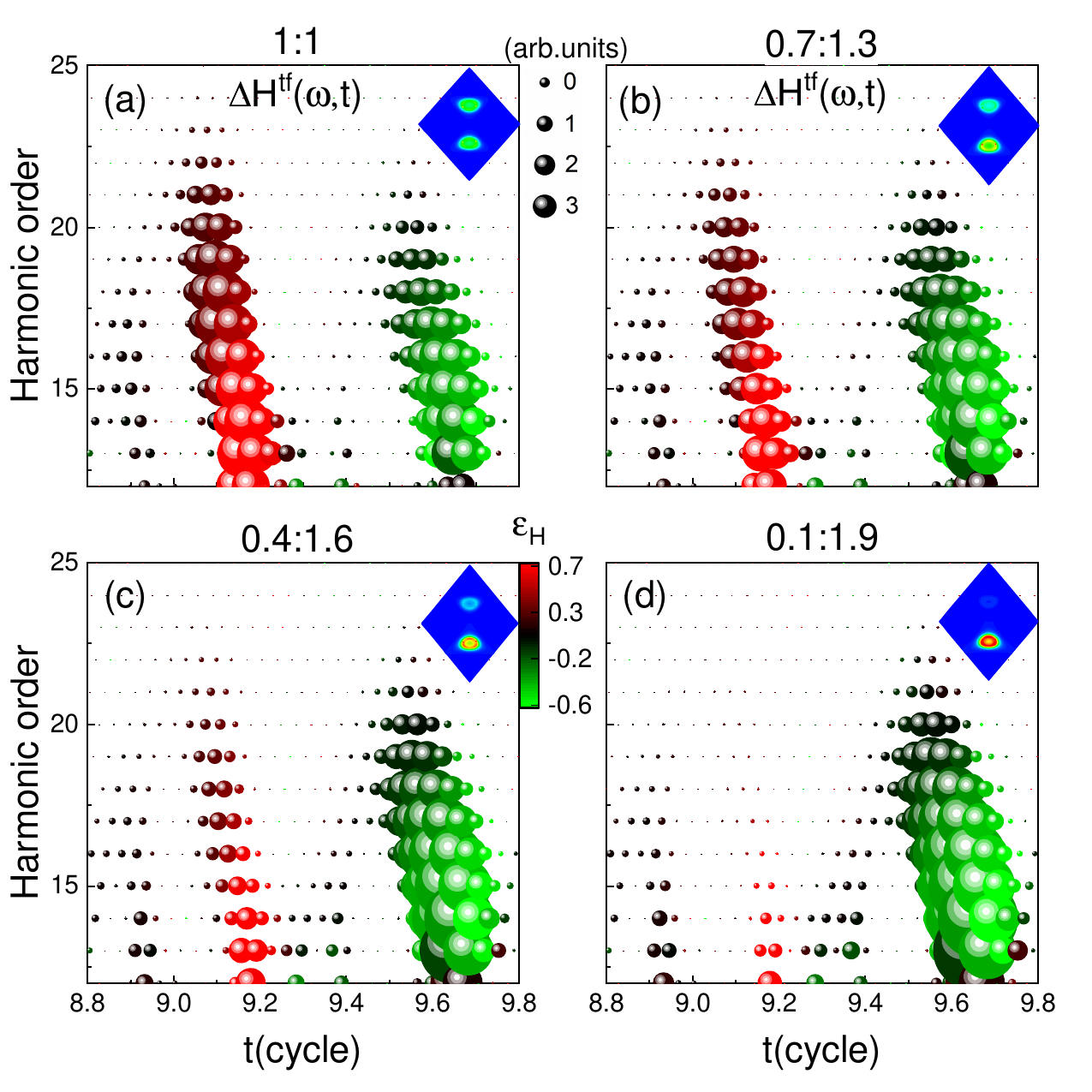}
\caption{
(a)–(d) Same as Fig. 5(b), but for different ratios of the $c$-band populations in the $\textrm{K}_{1}$ and $\textrm{K}_{1}$ valleys induced by the circularly polarized pulse, namely, $N_{\textrm{K}_{1}} : N_{\textrm{K}_{2}}= 1:1$, $0.7:1.3$, $0.4:1.6$, and $0.1:1.9$, respectively, as indicated in the insets.
}
\end{center}
\end{figure}

In our numerical simulations, varying the parameters of the circularly polarized pulse allows the valley polarization to be tuned only within a limited range.
To further investigate the influence of pre-excited electrons on the emission of left- and right-handed harmonics, we artificially vary the pre-excitation populations in the $\textrm{K}_1$ and $\textrm{K}_2$ valleys.
In the following, we denote the crystal momenta in the $\textrm{K}_1$ and $\textrm{K}_2$ valleys by $\mathbf{k}_1$ and $\mathbf{k}_2$, respectively. 
These momenta are related by inversion symmetry about the high-symmetry $\textrm{M}$ point.
The $c$-band populations in the $\textrm{K}_1$ and $\textrm{K}_2$ valleys induced by the pre-excitation pulse used in Fig. 5(a) are denoted by $\rho_{cc,0}^{\mathbf{k}_1}$ and $\rho_{cc,0}^{\mathbf{k}_2}$, respectively.
We take $\rho_{cc,0}^{\mathbf{k}_2}$ as a common reference and construct the initial $c$-band populations as $\rho_{cc}^{\mathbf{k}_1}=N_{\textrm{K}_1}\rho_{cc,0}^{\mathbf{k}_2}$
and
$\rho_{cc}^{\mathbf{k}_2}=N_{\textrm{K}_2}\rho_{cc,0}^{\mathbf{k}_2}$.
The corresponding $v$-band populations are given by $\rho_{vv}^{\mathbf{k}_1}=1-\rho_{cc}^{\mathbf{k}_1}$ and $\rho_{vv}^{\mathbf{k}_2}=1-\rho_{cc}^{\mathbf{k}_2}$.
These populations are then used as the initial conditions for the subsequent evolution driven by the elliptically polarized laser field. 
The corresponding results are shown in Fig. 6.

Building on the results in Fig. 5(b), Fig. 6 further demonstrates the dependence of the harmonic intensity difference $\Delta H^{\textrm{tf}} (\omega,t)$ on the valley polarization.
Here, the $c$-band population ratio between the $\textrm{K}_{1}$ and $\textrm{K}_{2}$ valleys is varied from $N_{\textrm{K}_{1}} : N_{\textrm{K}_{2}}= 1:1$ [Fig. 6(a)] to $0.7:1.3$ [Fig. 6(b)], $0.4:1.6$ [Fig. 6(c)], and $0.1:1.9$ [Fig. 6(d)], as indicated in the insets. 
As the population in the $\textrm{K}_{1}$ valley decreases while that in the $\textrm{K}_{2}$ valley increases, the intensity differences $\Delta H^{\textrm{tf}} (\omega,t)$ of the two polarized components exhibit opposite trends.
Specifically, the intensity differences $\Delta H^{\textrm{tf}} (\omega,t)$ for the right-circularly polarized component gradually decreases, whereas that for the left-circularly polarized component gradually increases. 
It can be found that the evolution of the harmonic intensity differences $\Delta H^{\textrm{tf}} (\omega,t)$ can reflect the variation of the conduction-band populations in the two valleys.

\begin{figure}[t]
\begin{center}
\includegraphics[width=8cm,height=7cm]{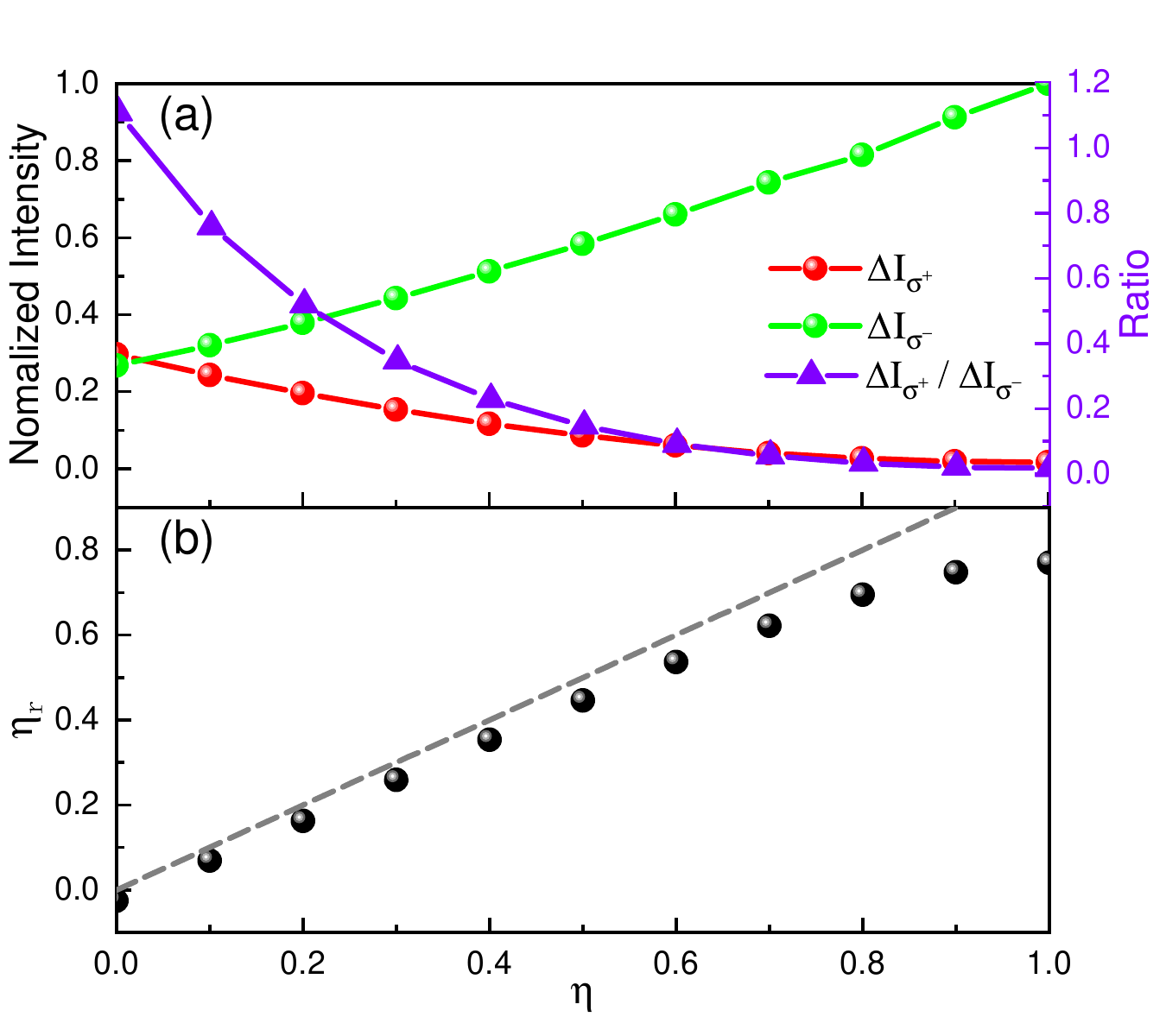}
\caption{
(a) Normalized intensities $\Delta I_{\sigma^{+}}$ and $\Delta I_{\sigma^{-}}$, and their intensity ratio $\Delta I_{\sigma^{+}}/ \Delta I_{\sigma^{-}}$ as functions of the valley polarization $\eta$.
(b) Reconstructed valley polarization $\eta_{r}$.
The gray dashed line indicates the ideal reconstruction.
}
\label{fig:graph1}
\end{center}
\end{figure}

\subsection{Reconstruction mechanism for valley polarization}

The above results suggest that the valley polarization can be reconstructed by establishing the relationship between $\Delta H^{\textrm{tf}} (\omega,t)$ and the pre-excited populations.
Here, the valley polarization is defined as $\eta = (N_{\textrm{K}_2}-N_{\textrm{K}_1}) / (N_{\textrm{K}_2}+N_{\textrm{K}_1})$, which is non-negative in our simulations because $N_{\textrm{K}_2} \geq N_{\textrm{K}_1}$.
To quantify the harmonic intensity changes, we define the integrated intensity differences of the right- and left-circularly polarized components, $\Delta I_{\sigma^{+}}$ and $\Delta I_{\sigma^{-}}$, respectively, as 
\begin{subequations}
\begin{align}
\Delta I_{\sigma^{+}} = \sum_{\omega}^{\omega \in [13,20]\omega_0} \dfrac{2}{T} \int_{8.8T}^{9.3T} \Delta H^{\textrm{tf}} (\omega,t) dt, \\
\Delta I_{\sigma^{-}} = \sum_{\omega}^{\omega \in [13,20]\omega_0} \dfrac{2}{T} \int_{9.3T}^{9.8T} \Delta H^{\textrm{tf}} (\omega,t) dt.
\end{align}
\end{subequations}
We assume that their ratio satisfies $\Delta I_{\sigma^{+}}/\Delta I_{\sigma^{-}} \propto N_{\textrm{K}_{1}}^2/ N_{\textrm{K}_{2}}^2$ (see Appendix B for a simple derivation). 
Therefore, the valley polarization can be reconstructed as
\begin{align}
\eta_r = \dfrac{1-\sqrt{\Delta I_{\sigma^{+}}/\Delta I_{\sigma^{-}}}}{1+\sqrt{\Delta I_{\sigma^{+}}/\Delta I_{\sigma^{-}}}}.
\end{align}

Figure 7(a) shows the normalized intensity differences $\Delta I_{\sigma^{+}}$ and $\Delta I_{\sigma^{-}}$ calculated from Eqs. (9a) and (9b), together with their ratio $\Delta I_{\sigma^{+}}/\Delta I_{\sigma^{-}}$ as functions of valley polarization $\eta$.
As $\eta$ increases, $\Delta I_{\sigma^{+}}$ decreases monotonically, whereas $\Delta I_{\sigma^{-}}$ increases monotonically. 
Accordingly, their intensity ratio $\Delta I_{\sigma^{+}}/\Delta I_{\sigma^{-}}$ also decreases monotonically.
This monotonic dependence allows the valley polarization to be reconstructed according to Eq. (10). 
The reconstructed valley polarization $\eta_r$ is shown in Fig. 7(b), which agrees well with the input polarization $\eta$, particularly in the low-polarization regime.
These results demonstrate the feasibility of quantitatively probing valley polarization through helicity-resolved harmonic emission.

\section{Conclusion}

\label{s5}

In summary, we finds that for the gapped graphene model, the two inequivalent valleys contribute harmonics with opposite helicities, and the elliptically polarized laser can provide temporal separation of these contributions within a single optical cycle.
The same half-cycle-dependent helicity switching is also observed in TDDFT calculations for monolayer MoS$_2$, indicating that this behavior is not limited to the gapped graphene model.
Importantly, valley polarization leaves a distinct signature in the helicity-resolved harmonic response through Pauli blocking. 
This signature enables the valley polarization to be extracted optically from the changes in harmonic intensities. 
Our work establishes chiral high-harmonic emission as a sensitive probe of valley populations and opens a route toward ultrafast, all-optical detection of valley dynamics in two-dimensional materials.

\section*{ACKNOWLEDGMENTS}

This work is supported by the National Natural Science Foundation of China (Grant No. 12404394 and No. 12347165), 
the Science and Technology Project of Hebei Education Department (Grant No. QN2026281), 
the Natural Science Foundation of Hebei Province, China (Grant No. A2026201012),
the Hebei University Natural Science Interdisciplinary Research Project (DXK202510), 
and the High-Performance Computing Center of Hebei University.

{\section*{DATA AVAILABILITY}

The data that support the findings of this article are not
publicly available. The data are available from the authors
upon reasonable request.

\appendix

\section{VALLEY-DEPENDENT CHIRAL HARMONIC EMISSION FOR DIFFERENT BAND GAPS}

\begin{figure}[t]
\begin{center}
\includegraphics[width=8.5cm,height=11cm]{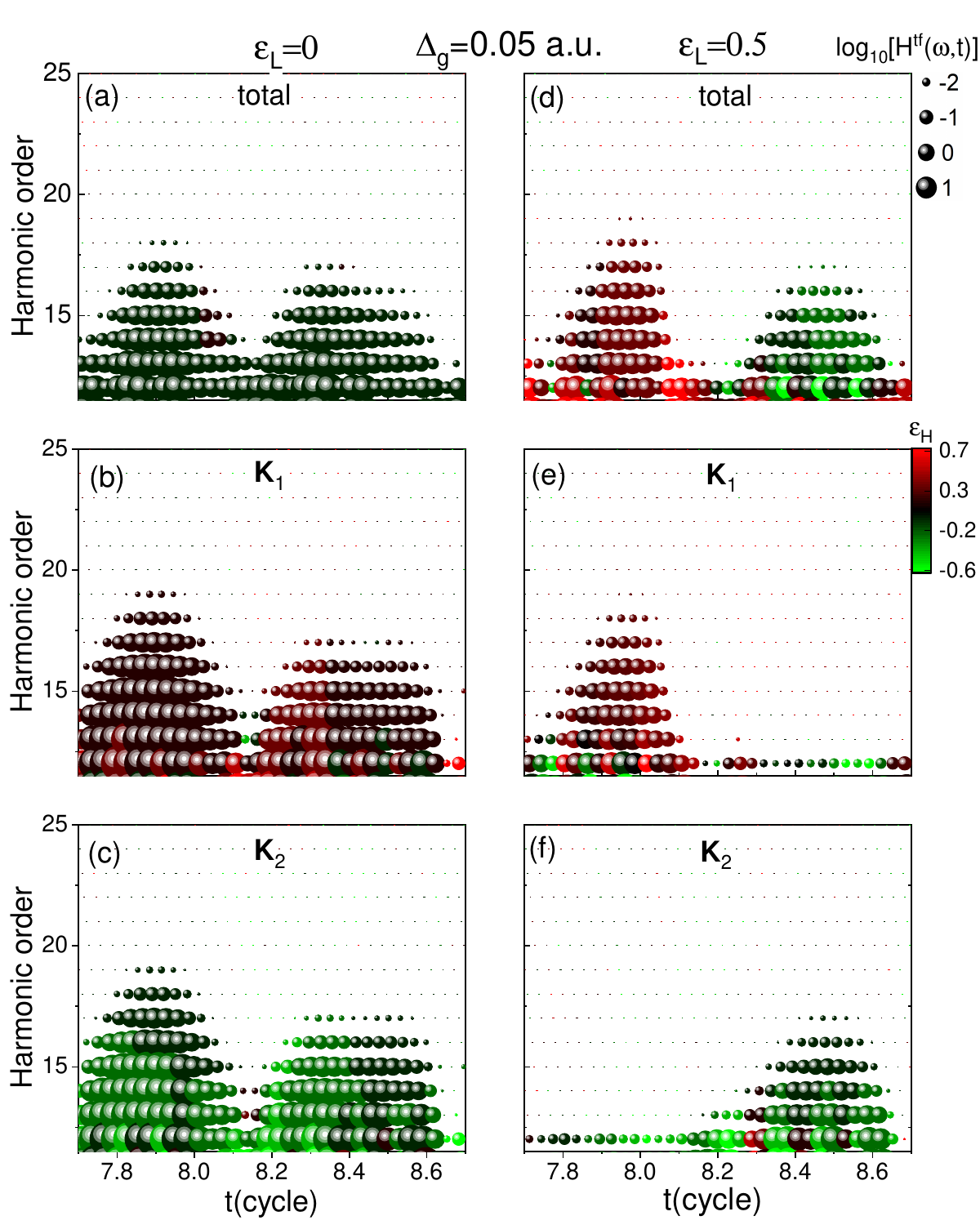}
\caption{
(a) Time-frequency distribution of the total harmonic current under linearly polarized ($\varepsilon_L=0$) for gapped graphene with $\Delta_g = 0.05$ a.u.
(b) (c) Valley-resolved time-frequency maps of the harmonic emission from the $\textrm{K}_1$ and $\textrm{K}_2$ valleys.
(d-f) Same as (a-c), but for elliptically polarized ($\varepsilon_L=0.5$) driving fields.
}
\label{fig:graph1}
\end{center}
\end{figure}

\begin{figure}[t]
\begin{center}
\includegraphics[width=8.5cm,height=11cm]{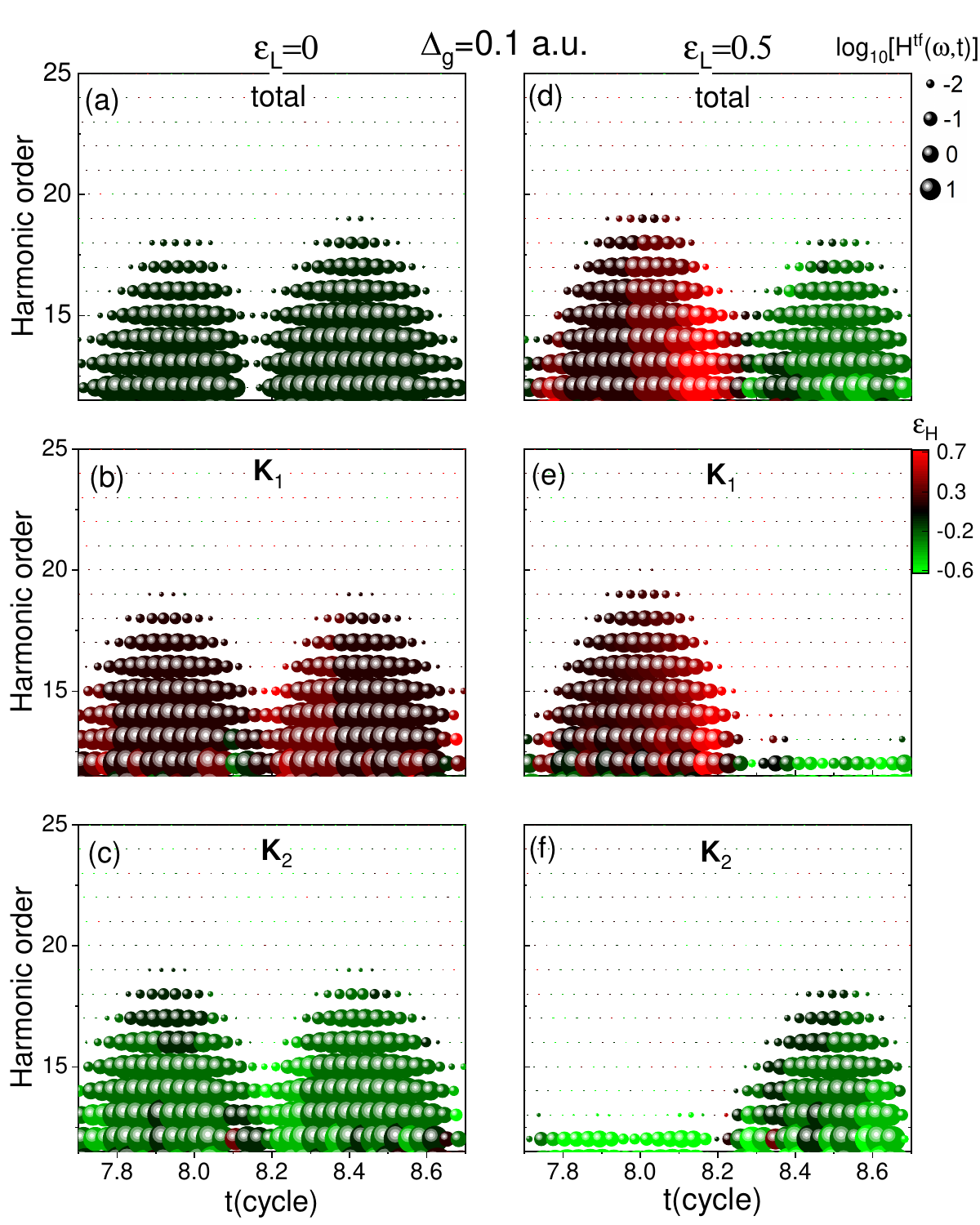}
\caption{Same as Fig. 8, but for a band gap of $\Delta_g=0.1$ a.u. 
}
\end{center}
\end{figure}

Here, we examine the robustness of the valley-dependent chiral harmonic emission for different band gaps.
Figures 8 and 9 show the time-frequency distributions of harmonic emissions from gapped graphene with $\Delta_g=0.05$ a.u. and $0.10$ a.u., respectively. 
For a linearly polarized driving field with $\varepsilon_L=0$, as shown in Figs. 8(a)–8(c) and 9(a)–9(c), the total harmonic emission [Figs. 8(a) and 9(a)] has nearly zero ellipticity. 
In contrast, the $\textrm{K}_1$ valley emits right-handed harmonics, whereas the $\textrm{K}_2$ valley emits left-handed harmonics [Figs. 8(b), 8(c), 9(b), and 9(c)].

When the laser ellipticity is increased to $\varepsilon_L=0.5$, the total harmonic emission exhibits a pronounced temporal asymmetry. 
As shown in Figs. 8(d) and 9(d), intense right-handed harmonics are emitted during the first half of the optical cycle, whereas weaker left-handed harmonics appear during the second half. 
The valley-resolved results reveal the origin of this behavior. 
The $\textrm{K}_1$ valley mainly contributes in the first half cycle and generates intense right-handed harmonics [Figs. 8(e) and 9(e)]. 
Whereas its contribution is strongly suppressed in the second half cycle. 
In contrast, the $\textrm{K}_2$ valley is suppressed during the first half cycle and becomes dominant in the second half cycle, where it generates strong left-handed harmonics [Figs. 8(f) and 9(f)].

These results are consistent with those obtained for $\Delta_g=0.07$ a.u. in the main text. 
They show that the valley-dependent chiral harmonic emission and its temporal separation are robust over a wide range of band gaps.

\section{DERIVATION FOR THE RELATION BETWEEN CHIRAL HARMONICS AND THE VALLEY POLARIZATION}

For a lattice momentum $\mathbf{k}_1$ in the $\textrm{K}_1$ valley, according to Eq. (1), one can obtain $\frac{d}{d t} \rho_{cv}^{\mathbf{k}_{1}}(t) \propto \rho_{vv}^{\mathbf{k}_{1}}(t) - \rho_{cc}^{\mathbf{k}_{1}}(t)$. 
In the absence of pre-excitation by the circularly polarized laser, under the strong-field approximation, both $\rho_{cv}^{\mathbf{k}_{1}}(t)$ and the corresponding time-frequency current $\textbf{\textit{j}}_{\mathbf{k}_{1}}^{\textrm{tf}}(\omega,t)$ proportional to this population difference $\rho_{vv}^{\mathbf{k}_{1}}(t) - \rho_{cc}^{\mathbf{k}_{1}}(t) \approx 1$. 
In contrast, a circularly polarized pulse can populate the $c$ band before the arrival of the elliptically polarized field.
Here, we use $\rho_{cc}^{\mathbf{k}_1}$ to denote the residual $c$-band population after the pulse. 
The subsequent interband excitation is then suppressed by Pauli blocking.
The population difference can be approximated as $\rho_{vv}^{\mathbf{k}_{1}}(t) - \rho_{cc}^{\mathbf{k}_{1}}(t) \approx (1-2\rho^{\mathbf{k}_{1}}_{cc})$.
It follows that $\frac{d}{d t} \rho_{cv}^{\mathbf{k}_{1},\circlearrowleft}(t)  \propto (1-2\rho^{\mathbf{k}_{1}}_{cc})$,
and thus $\textbf{\textit{j}}_{\mathbf{k}_{1},\circlearrowleft}^{\textrm{tf}}(\omega,t) \propto (1-2\rho^{\mathbf{k}_{1}}_{cc})$.
The current difference between the cases without and with pre-excitation is therefore $\Delta \textbf{\textit{j}}_{\mathbf{k}_{1}}^{\textrm{tf}}(\omega,t) = \textbf{\textit{j}}_{\mathbf{k}_{1}}^{\textrm{tf}}(\omega,t) - \textbf{\textit{j}}_{\mathbf{k}_{1},\circlearrowleft}^{\textrm{tf}}(\omega,t) \propto 2 \rho^{\mathbf{k}_{1}}_{cc}$.
Consequently, the change in the $\sigma^{+}$ harmonic is $\Delta I_{\sigma^{+}}  \propto \Delta H^{\textrm{tf}}_{\textrm{K}_{1}} (\omega,t) = \left| \sum_{\mathbf{k}_{1}} \Delta \textbf{\textit{j}}_{\mathbf{k}_{1}}^{\textrm{tf}}(\omega,t)  \right|^2 \propto 4 \left| \sum_{\mathbf{k}_{1}}  \rho^{\mathbf{k}_{1}}_{cc} \right|^2 = 4 N_{\textrm{K}_1}^2$.

The same argument applies to the lattice momentum $\mathbf{k}_2$ in the $\textrm{K}_2$ valley. 
(Here, $\mathbf{k}_1$ and $\mathbf{k}_2$ are related by inversion symmetry about the high-symmetry $\textrm{M}$ point.) 
One can obtain $\Delta I_{\sigma^{-}}  \propto \Delta H^{\textrm{tf}}_{\textrm{K}_{2}} (\omega,t)  \propto  4 N_{\textrm{K}_2}^2$.
Combining the results for the two valleys gives $\Delta I_{\sigma^{+}} / \Delta I_{\sigma^{-}} \propto N_{\textrm{K}_{1}}^2/ N_{\textrm{K}_{2}}^2$.

\end{document}